\documentclass{aastex}
\usepackage{apj}
\shorttitle{DIRECT Distances to Galaxies. VI}
\shortauthors{Macri {\it et al.}}

\def \nd  {\nodata}
\def \nto {1298}
\def \ntr {780}

\def \nte {46}
\def \ntp {26}
\def \ntm {169}

\def \ngd {544}
\def \nac {251}
\def \nae {47}
\def \nap {62}
\def \nam {184}

\def \de {^{\circ}}
\def \hr {^{\rm h}}
\def \mi {^{\rm m}}
\def \se {^{\rm s}}

\def \vi {V\!-\!I}
\def \bv {B\!-\!V}

\begin{document}
\title{DIRECT Distances to Nearby Galaxies Using Detached Eclipsing Binaries
and Cepheids. VI. Variables in the Central Part of M33\altaffilmark{1}}

\author{L.M.~Macri, K.Z.~Stanek\altaffilmark{2}, D.D.~Sasselov\altaffilmark{3}
\& M.~Krockenberger}
\affil{Harvard-Smithsonian Center for Astrophysics, 60 Garden St., Cambridge MA
02138, USA}
\email{lmacri, kstanek, sasselov, krocken@cfa.harvard.edu}

\and

\author{J.~Kaluzny}
\affil{N. Copernicus Astronomical Center, Bartycka 18, PL-00-716 Warszawa,
Poland}
\email{jka@camk.edu.pl}

\altaffiltext{1}{Based on observations collected at the Fred L. Whipple
Observatory 1.2-m telescope and at the Michigan-Dartmouth-MIT 1.3-m telescope.
\baselineskip=18pt}
\altaffiltext{2}{Hubble Fellow}
\altaffiltext{3}{Alfred P. Sloan Foundation Fellow}

\begin{abstract}

The DIRECT project aims to determine direct distances to two important galaxies
in the cosmological distance ladder -- M31 and M33 -- using detached eclipsing
binaries (DEBs) and Cepheids. 

We present the results of the first large-scale CCD-based search for variables
in M33. We have observed two fields located in the central region of M33 for a
total of 95 nights on the F. L. Whipple Observatory 1.2~m telescope and 36
nights on the Michigan-Dartmouth-MIT 1.3~m telescope. We have found a total of
\ngd\ variables, including \nac\ Cepheids and \nae\ eclipsing binaries.

The catalog of variables is available online, along with finding charts and BVI
light curve data (consisting of $8.2\times 10^4$ individual measurements). The
complete set of CCD frames is available upon request.

\end{abstract}


\vspace{-12pt}

\section{Introduction}

\vspace{-2.3pt}

The DIRECT project (as in ``direct distances'') started in 1996 with the
long-term goal of obtaining distances to two important galaxies in the
cosmological distance ladder -- M31 and M33 -- using detached eclipsing
binaries (DEBs) and Cepheids.  These two nearby galaxies are the stepping
stones in most of the current effort to understand the evolving universe at
large scales. Not only are they essential to the calibration of the
extragalactic distance scale, but they also constrain population synthesis
models for early galaxy formation and evolution. However, accurate distances
are essential to make these calibrations free from large systematic
uncertainties.

Detached eclipsing binaries have the potential to establish distances to M31
and M33 with an unprecedented accuracy of better than 5\% and possibly to
better than 1\%. Current uncertainties in the distances to these galaxies are
in the order of 10 to 15\%, as there are discrepancies of 0.2-0.3 mag between
various distance indicators. Detached eclipsing binaries \citep{an91,pa97}
offer a single-step distance determination to nearby galaxies and may therefore
provide an accurate zeropoint calibration for other distance indicators,
including Cepheids. This work is the first paper in the DIRECT series related
to the discovery of variables in M33. The previous papers in the series
\citep{ka98, st98, st99, ka99, mo99} have presented results for five fields in
M31.

This work represents the first large-scale CCD-based search for variables in
M33. A previous paper \citep{ma00a} presented the details of data acquisition,
reduction and calibration, as well as positions and mean BVI magnitudes, for
57,581 objects in the central part of M33 (consisting of the DIRECT M33 Survey
fields A, B and C). The present paper presents the analysis of the variable
stars found in fields A and B. \citet{st01} will present the analysis of the
variable stars found in field C, which benefits from an additional year of
observations. 

\S 2 summarizes the details of data acquisition and reduction; \S 3 discusses
the search and classification of variable stars, and presents the catalogs of
variable stars; \S4 contains a discussion of our results; and \S5 compares our
sample of variables with previous catalogs.

\vspace{-6pt}

\section{Observations and data reduction}

M33 was observed at the Fred L. Whipple Observatory 1.2-m telescope and at the
Michigan-Dartmouth-MIT 1.3-m telescope between September 1996 and October
1997. Useful data was acquired on 42 nights during the program. We observed
three fields located north, south and south-west of the center of M33, which we
labeled A, B and C, respectively. This paper presents the analysis of the
variable stars located in fields A and B, whose J2000.0 center coordinates are
R.A. = $01\hr 34\mi 05.1\se$, Dec.= $30\de 43\mi 43\se$ and R.A. = $01\hr 33\mi
55.9\se$, Dec.= $30\de 34\mi 04\se$, respectively.

\citet{ma00a} contains the details of data acquisition, reduction and
calibration, as well as a catalog of positions and mean BVI magnitudes of the
objects present in these fields. We refer the interested reader to that
publication for details. PSF photometry was carried out following the precepts
of \citet{ka98}, using an automated pipeline and the DAOPHOT and ALLSTAR
programs \citep{st87,st92} . The photometric calibration of the fields was
carried out using standard stars from \citet{la92}; the uncertainty in our
zeropoints is $\pm0.04$~mag.

\section{Variable search and classification}

\citet{ka98} presented a detailed explanation of our techniques for period
determination and variable classification, which we summarize here. The reader
should also note that our classification criteria for variables have been
slightly modified over the past five papers in the series.

In \citet {ma00a}, we calculated the $J_S$ variability index \citep{st96} for
all objects in our database, using the $V$-band data since it has the best
sampling.  We flagged objects with $J_{S,V} \geq 0.75$ and $\sigma_V \geq
0.04$~mag as candidate variables.  The second criterion was introduced to
reject bright stars with very small-amplitude variability. This yielded \nto\
candidate variables out of a total of 40,642 stars.

Periods were determined using the $V$-band data only, employing a variant of
the Lafler-Kinman technique proposed by \citet{st96}. We searched for periods
between 0.25 and 400 days, which corresponds to the time span of the
observations. When fitting a light curve to a Cepheid or eclipsing binary
template, we tried the ten most likely periods derived by this procedure, as
well as periods corresponding to one-half of their values. Once the best-fit
period was identified, it was refined by trying ten values in its close
vicinity.

Figure 1 shows our phase coverage as a function of period. The ten most
likely periods of each object were saved for the next step in the analysis
procedure, involving the classification of variables. We classified variables
into four categories: Cepheids, eclipsing binaries, periodic and
miscellaneous. The classification process involved automated template fits as
well as a visual examination of the phased light curves and of the CCD image of
every object.

A variable star was classified as a Cepheid if the $\chi_\nu^2$ of a fit of its
time series to a template Cepheid light curve was smaller than the $\chi_\nu^2$
of a fit to a template EB light curve and 3 times smaller than the $\chi_\nu^2$
of a fit to a straight line.  We further required that the $V$-band amplitude
be larger than 0.1~mag. We used the parameterization of Cepheid light curves in
the $V$ and $I$ bands from \citet{st96}. Since $B$-band templates were not
available, we used the $V$-band ones, with their amplitudes rescaled by a fixed
factor of 1.45. Mean magnitudes were obtained by numerical integration of the
best-fit templates.

A variable star was classified as an eclipsing binary (EB) if the $\chi_\nu^2$
of a fit of its time series to a EB light curve was smaller than the
$\chi_\nu^2$ of a fit to a template Cepheid light curve and 1.75 times smaller
than the $\chi_\nu^2$ of a fit to a straight line.  The EB light curve was
modeled using nine parameters: the period ($P$), the zero point of the phase
($T_0$), the eccentricity ($e$), the longitude of periastron ($\omega$), the
radii of the two stars relative to the binary separation ($r_1$ and $r_2$), the
inclination angle ($i$), and the magnitudes of the primary and of the
uneclipsed system. We further required that the larger stellar radius be less
than 90\% of the binary separation and that the light of the brighter star be
less than 90\% of the total light. Lastly, we rejected periods within 0.025d of
1 and 2 days, to avoid spurious classifications due to aliasing of
long-period variables. \nte\ objects were classified as EBs.

In some cases, the classification routine was able to fit a variable light
curve with either a Cepheid or EB template. However, visual inspection of the
light curve showed unusual colors or a shallow amplitude. These objects may be
Cepheids or EBs, but higher-quality observations are required for a secure
classification.  Therefore, we have only classified them as ``periodic''.
\ntp\ objects fell into this category.

In 206 cases, the classification routine could not obtain a satisfactory match
to a Cepheid or EB template, but the $V$-band $J_{S}$ index of the object being
analyzed was equal or greater than 1.2, implying a well-established
variability. We examined the light curve and CCD image of these objects and
rejected 37 of them because they were close to or on top of chip defects, or
near saturated stars. The remaining \ntm\ objects seemed to be authentic and
thus were classified as ``miscellaneous''.  An additional \ntr\ objects, which
did not match the Cepheid or EB templates and which had a $V$-band $J_{S} < 1.2$,
were discarded.

The star catalog of \citet{ma00a} required that objects be present in at least
two bands (the $V$ band and either the $B$ or $I$ bands) in order to be
included in it.  However, there could be additional variables in our fields with data in
one band only.  This situation could arise in regions of high extinction, where
only the $I$-band flux would be detected. Additionally, some objects could have
been detected in the $V$ band and missed in the $I$ band due to close proximity to
a bright red object. Therefore, we decided to repeat our variable search and
classification for single-band objects in our $V$- and $I$-band databases.

We searched the $V$-band database for objects with no counterparts in other bands
and with $J_{S} \geq 0.75$ and $\sigma_V \geq 0.04$~mag.  We found 255 such
objects and analyzed them in manner described above.  11 were classified as
Cepheids, 1 was classified as an EB, 15 were labeled as ``miscellaneous'', and
228 were rejected.

We searched the $I$-band database for objects with no counterparts in other bands
and with $J_{S} \geq 1.2$ and $\sigma_I \geq 0.04$~mag. We were aware that
our observing strategy had not included obtaining back-to-back images in this
band, a fact that affects the meaning of the $J_S$ statistic. Therefore, a
substantial number of objects (638) passed our selection criteria. We decided
to only retain the variables that fitted a Cepheid or EB template. Since the
latter are blue objects, it is not surprising that none were found. 36 objects
had good fits to the Cepheid templates, but their location in the P-L plane
indicates that they are most likely Population II Cepheids (see \S4). We
therefore classified them as ``periodic''.

In summary, our classification of candidate variables resulted in the
selection of \ngd\ objects, of which \nac\ were classified as Cepheids, \nae\
as eclipsing binaries, \nap\ as periodic variables and \nam\ as miscellaneous
variables. The candidate objects include 11 Cepheids, 1 EB and 15
miscellaneous variables detected only in the $V$ band, and 36 periodic variables
detected only in the $I$ band. Figures 2-5 contain representative light curves
of each class of variables.

Tables 1-4 contain catalogs of the Cepheids, EBs, periodic and miscellaneous
variables, respectively, discovered in our search. We list their designations,
positions (in J2000.0 coordinates), periods ($J_{S}$ values for miscellaneous
variables), mean $V$, I, and $B$ magnitudes, and {\it r.m.s.} deviations of the
light curves from the best-fit templates (for Cepheids and EBs) or from the
mean magnitudes (for miscellaneous and periodic variables). Tables 5-8 contain
the individual photometric measurements for all variables\footnote{Also 
available through anonymous FTP on cfa-ftp.harvard.edu, /pub/kstanek/DIRECT.}.

\section{Discussion}

Figure 6 shows $(\bv,V)$ and $(\vi,V)$ color-magnitude diagrams indicating the
location of the variables which have data in more than one band. The star
catalog from \citet{ma00a} is plotted with black dots, while the variables are
marked with circles of different colors: green for Cepheids, blue for EBs, red
for miscellaneous variables, and yellow for periodic variables.  The
displacement induced by an extinction of $A_V = 1$~mag is indicated by the
arrows.

The segregation of EBs in the blue part of the diagram, and of Cepheids in the
instability strip, is quite clear in both panels. Variables labeled as
``miscellaneous'' seem to fall into two well-defined categories: i) luminous
blue variables ($V < 18, \vi < 1.5$) and ii) AGB stars ($V < 20.5, \vi > 2$),
while other miscellaneous variables populate the lower regions of the CMD ($V >
21, \vi \sim 2$).

Figure 7 shows the $B$, $V$ and $I$-band Period-Luminosity relations for the Cepheid
and ``periodic'' variables, constructed from the data presented in Table 1.
The $I$-band magnitudes of the long-period ``periodic'' variables are well
correlated, indicating that these are probably Population II variables (either
W Virginis or RV Tauri stars). The Cepheids for which HST archival data is
available are presently being studied by \citet{mo00}. The analysis of the
Cepheid sample and a determination of the distance to M33 will be presented in
\citet{ma00b}.

\section{Previously-known variables in M33}

The search for variables in M33 has a long and venerable history. \citet{du22}
(hereafter, Du22) discovered three variables in this galaxy while searching
for novae.  The first two of these are irregular variables, while the third
one turned out to be a 42-day Cepheid. \citet{hu26} (hereafter Hu26) undertook
an extensive observational program which resulted in the discovery of 41
variables, including 35 Cepheids and one eclipsing variable. Later,
\citet{hs53} (hereafter, HS53) discovered an additional three bright irregular
variables. After a long hiatus, \citet{vhk75} (hereafter VHK75) discovered 36
new variables but were unable to determine periods and classifications for
them, due to the rather poor quality of their magnitudes. Lastly, \citet{sc83}
(hereafter, SC83) discovered an additional 12 Cepheids.

Table 9 lists the properties of all these ``classical'' variables, and their
DIRECT counterparts, where available. Coordinates are from the catalog of
\citet{ma00a}, if the variable was present in it. Otherwise, they were
obtained from the original publication or from a digitized POSS image of M33.
We hope this table, which to our knowledge is the first compilation of all M33
classical variables, will serve as reference for future work.

In general, there is good agreement between the classical and the DIRECT
periods and classification of these variables. In two cases (Hu26-V04 and
VHK75-76), we believe the original finding charts may be erroneously
indicating a nearby star as the variable instead of the actual object. In both
cases, the star marked as variable in the original finding charts exhibits no
variability in our database, while a nearby star has the properties of the
object listed as variable in the text or tables of the original work. It seems
that \citet{vhk75} did not notice this mis-identification of V04 by
\citet{hu26}, and thought that VHK75-49 was a new object when it was actually
the same variable.

Four classical variables which had been not been classified as Cepheids were
found to be so by our automated template fitting program. Such was the case for
Du22-V01 (re-classified as a 33-day Cepheid), Hu26-V15 (re-classified as a
50-day Cepheid), Hu26-V21 (re-classified as a 67-day Cepheid) and Hu26-V45
(re-classified as a 74-day Cepheid). One classical Cepheid (Hu26-V19) is
present in the star catalog of \citet{ma00a}, but it did not meet the
criteria to be classified as a variable. Indeed, the amplitude of its
pulsation seems to have decreased from $\sim1$~mag at the beginning of the
century to a few hundredths of a magnitude at present. We are exploring this
change and will report our findings elsewhere.

The only modern search for variables in M33 was undertaken by \citet{kmw87}
(hereafter, KMW87).  The observations were carried out using plates, but these
were scanned and digitized using a PDS machine, and the search for variables
was performed using an automated software routine. They discovered 65 Cepheids,
44 long-period variables, 6 periodic variables and 224 miscellaneous
variables over a large area of the disk of M33.

Table 10 lists DIRECT counterparts for the KMW87 variables located in our
fields A \& B. Generally, there is good agreement between the KMW87 and DIRECT
periods and classification of these variables. We were able to retrieve all but
one of the KMW87 Cepheids; the lone exception seems to have a very faint
counterpart in our frames, below the detection threshold. We were also able to
recover the only KMW87 W Vir candidate located in our fields, which we have
classified as a short-period Cepheid, and all but three of the KMW87
long-period variables. Lastly, we found good matches ($<1\arcsec$) in our
catalog of variables for one third (23/72) of the KMW87 unclassified variables
present in our fields.

\section{Conclusions}

We have presented the results of the first large-scale CCD-based search for
variables in M33. Our catalog contains \ngd\ variables, including \nac\
Cepheids, \nae\ eclipsing binaries, \nap\ periodic variables, and \nam\
miscellaneous variables.

The Cepheids for which HST archival data is available are presently being
studied by \citet{mo00}. The analysis of the Cepheid sample and a determination
of the distance to M33 will be presented in \citet{ma00b}.

Finding charts and individual photometric measurements for all variables are
available from the DIRECT project Web site at {\tt
http://cfa-www.harvard.edu/ $\sim$kstanek/DIRECT}.

We would like to thank the telescope allocation committees of the FLWO and MDM
Observatories for the generous amounts of telescope time devoted to this
project. LMM would like to thank John Huchra for his support and comments. KZS
was supported by a Hubble Fellowship grant HF-01124.01-99A from the Space
Telescope Science Institute, which is operated by the Association of
Universities for Research in Astronomy, Inc., under NASA contract
NAS5-26555. DDS acknowledges support from the Alfred P. Sloan Foundation and
from NSF grant No. AST-9970812. JK was supported by KBN grant 2P03D003.17

\clearpage

\begin{figure}
\plotone{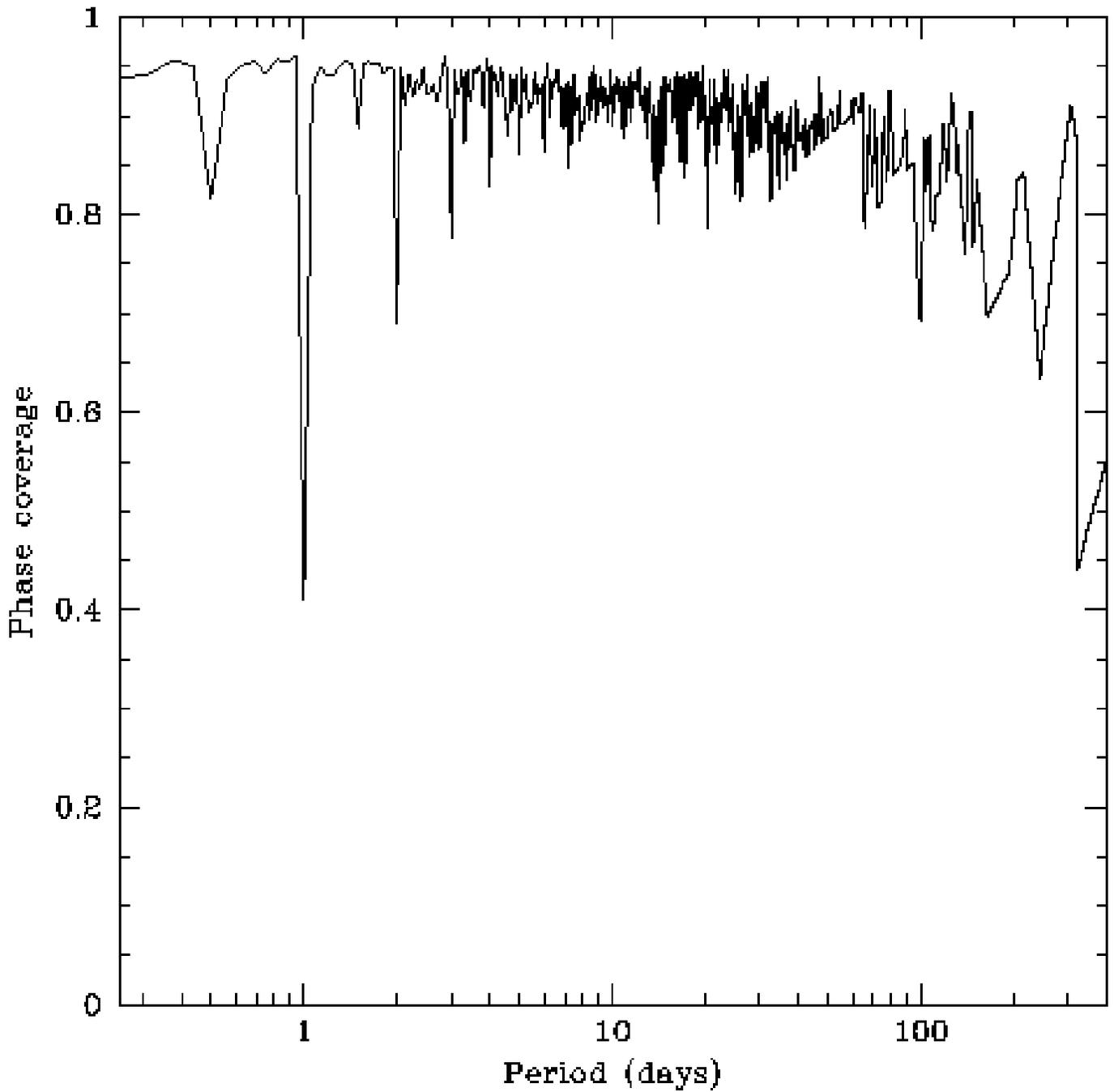}
\caption{Phase coverage for variables of different periods, based on the span
and sampling of our $V$-band data.}
\end{figure}

\clearpage

\begin{figure}
\plotone{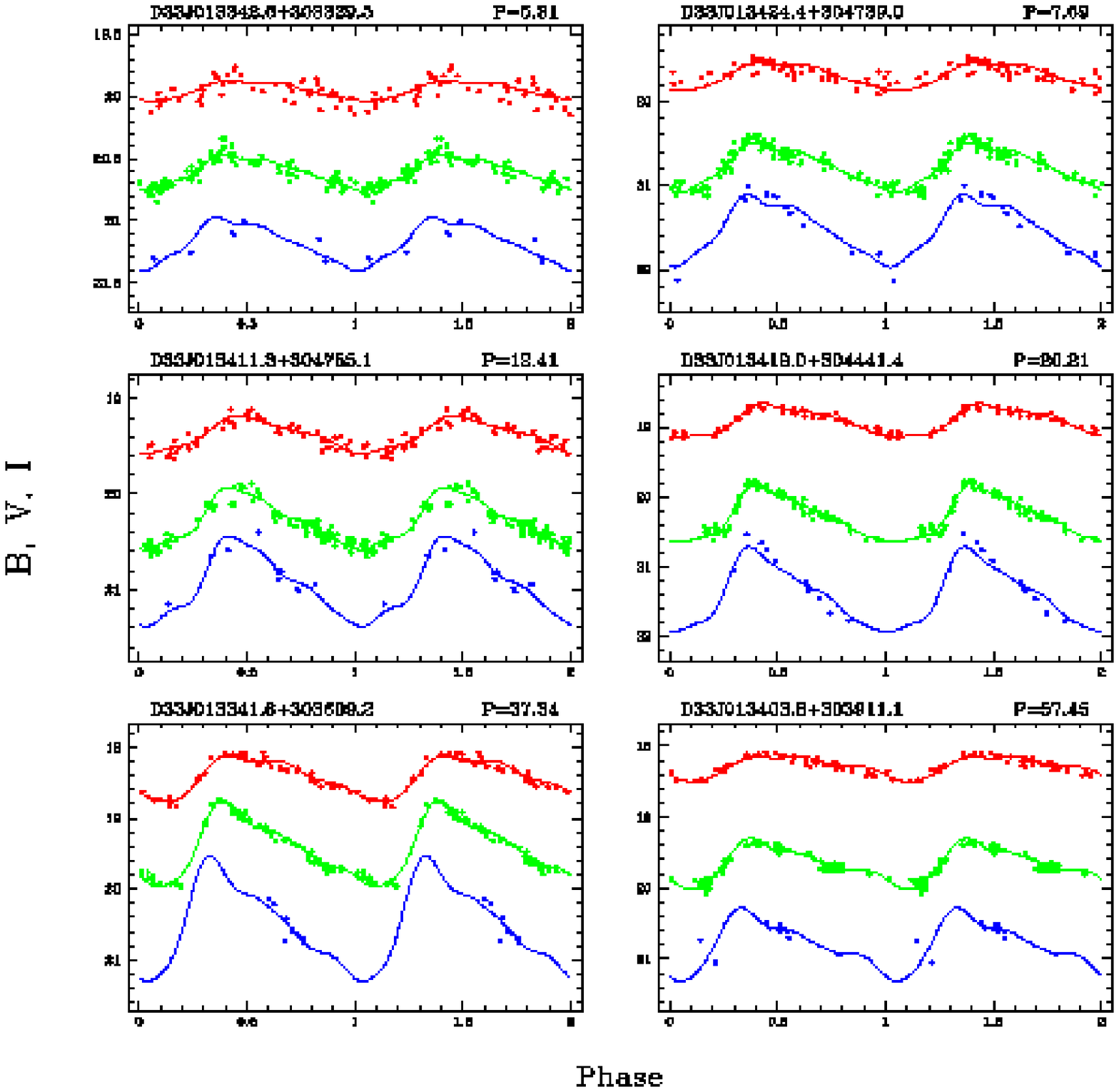}
\caption{Light Curves for 6 of the \nac\ Cepheid variables. $BVI$ data are
represented by blue, green and red filled circles, respectively. The best-fit
model light curves are overplotted using solid lines.}
\end{figure}

\clearpage

\begin{figure}
\plotone{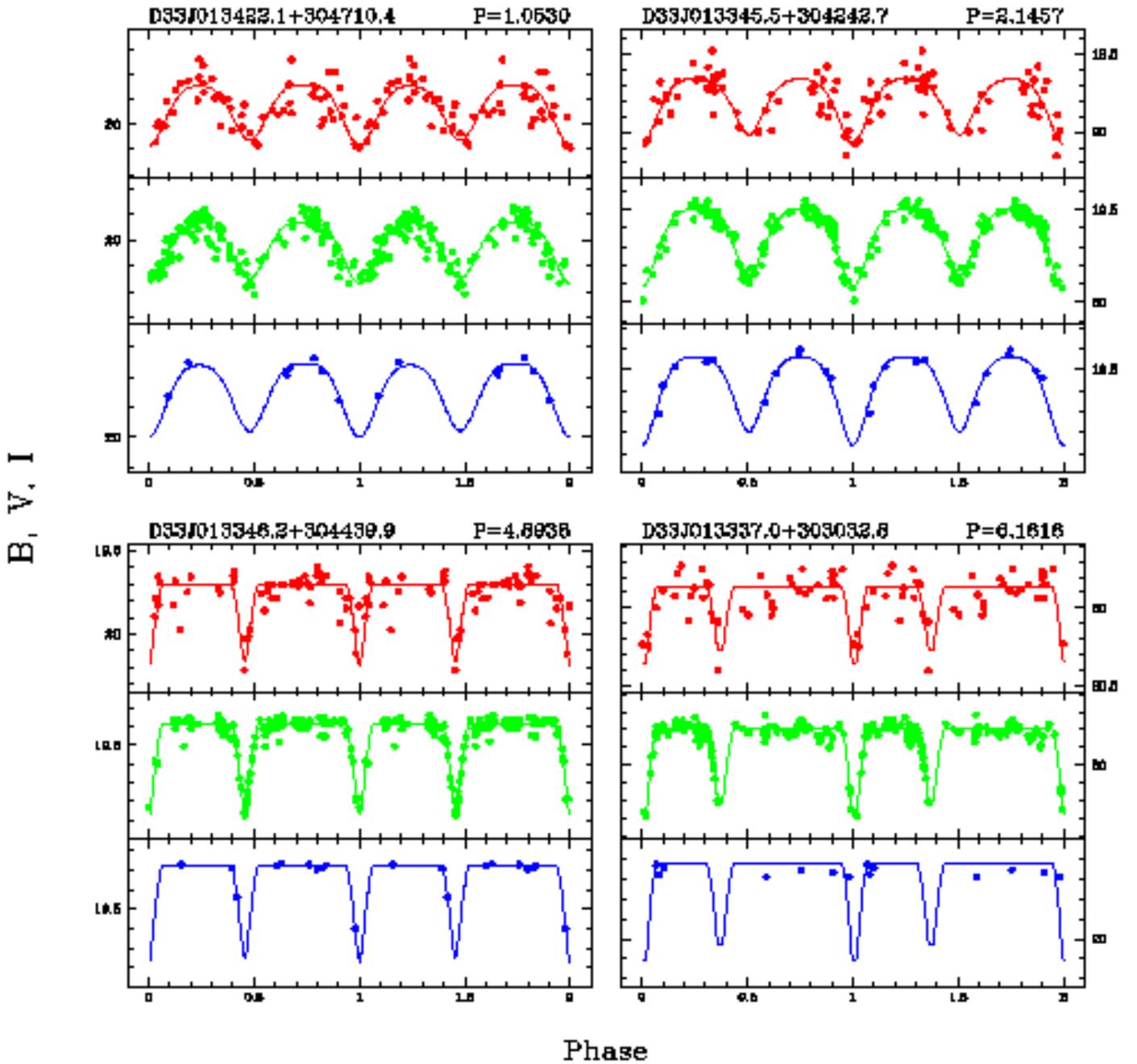}
\caption{Light Curves for 4 of the \nae\ eclipsing binaries. $BVI$ data are
represented by blue, green and red filled circles, respectively. The best-fit
model light curves are overplotted using solid lines.}
\end{figure}

\clearpage

\begin{figure}
\plotone{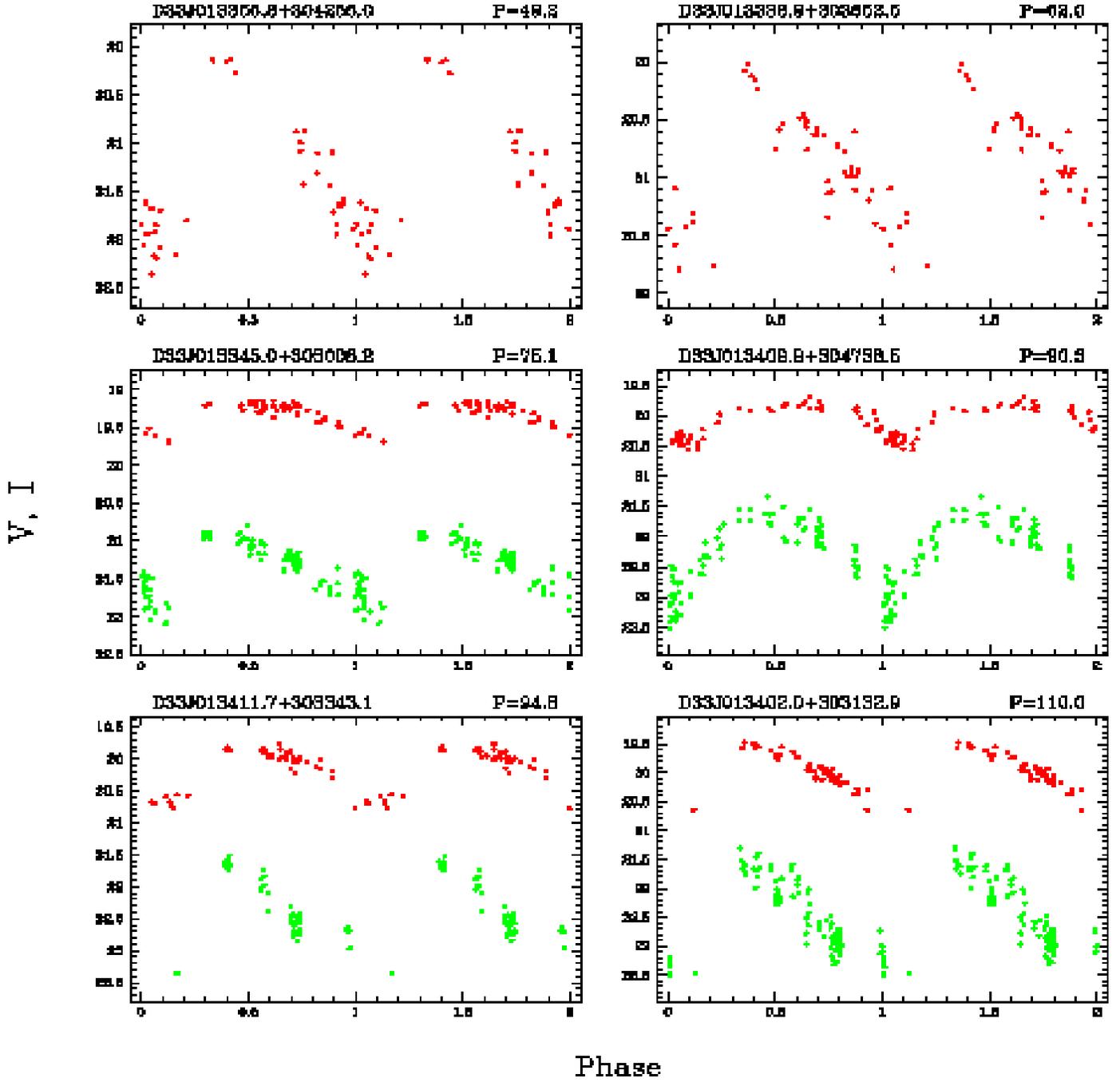}
\caption{Light Curves for 6 of the \nap\ periodic variables. $VI$ data are
represented by green and red filled circles, respectively.}
\end{figure}

\clearpage

\begin{figure}
\plotone{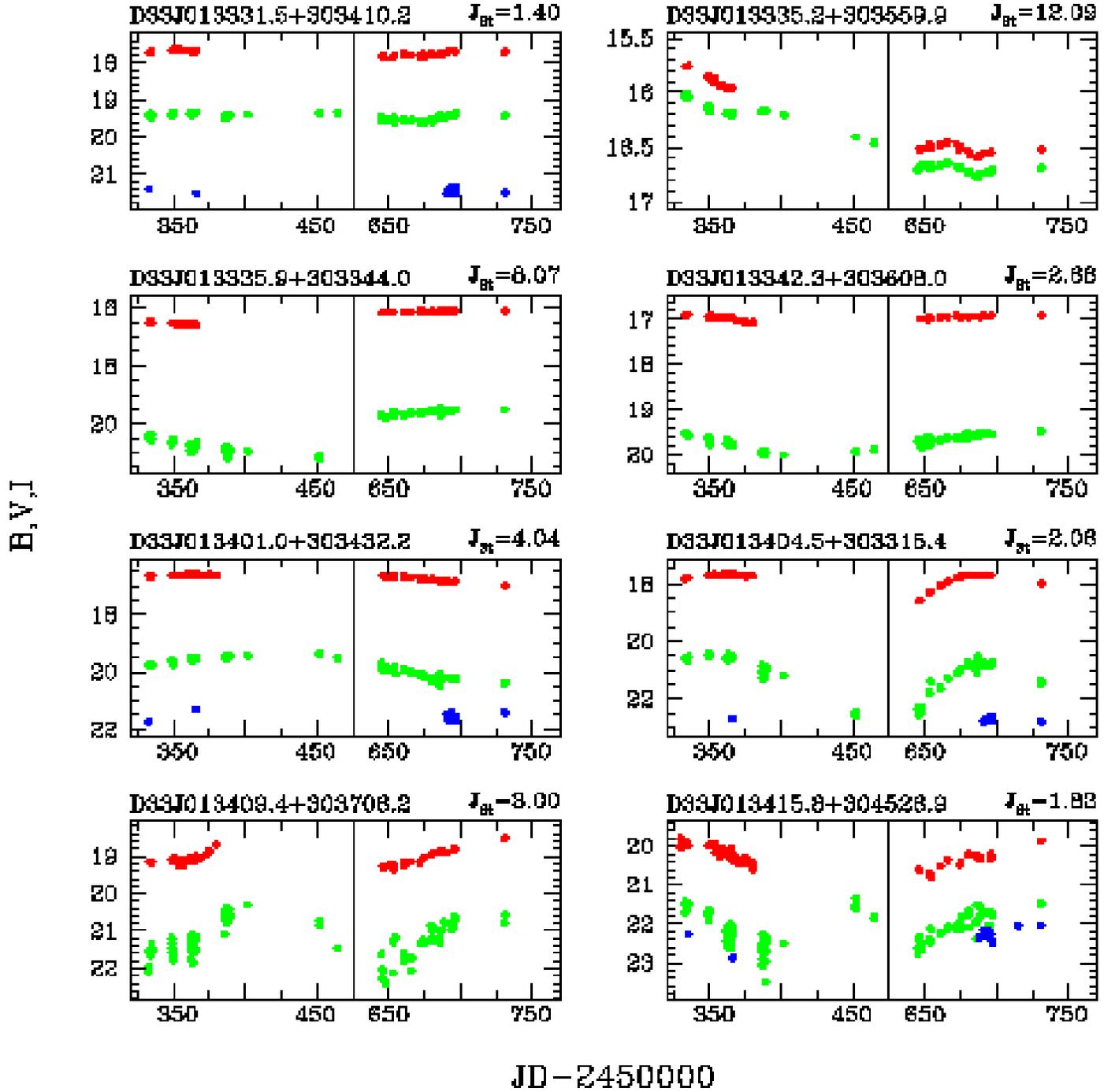}
\caption{Light Curves for 8 of the \nam\ miscellaneous variables. $BVI$ data
are represented by blue, green and red filled circles, respectively.}
\end{figure}

\clearpage

\begin{figure}
\plotone{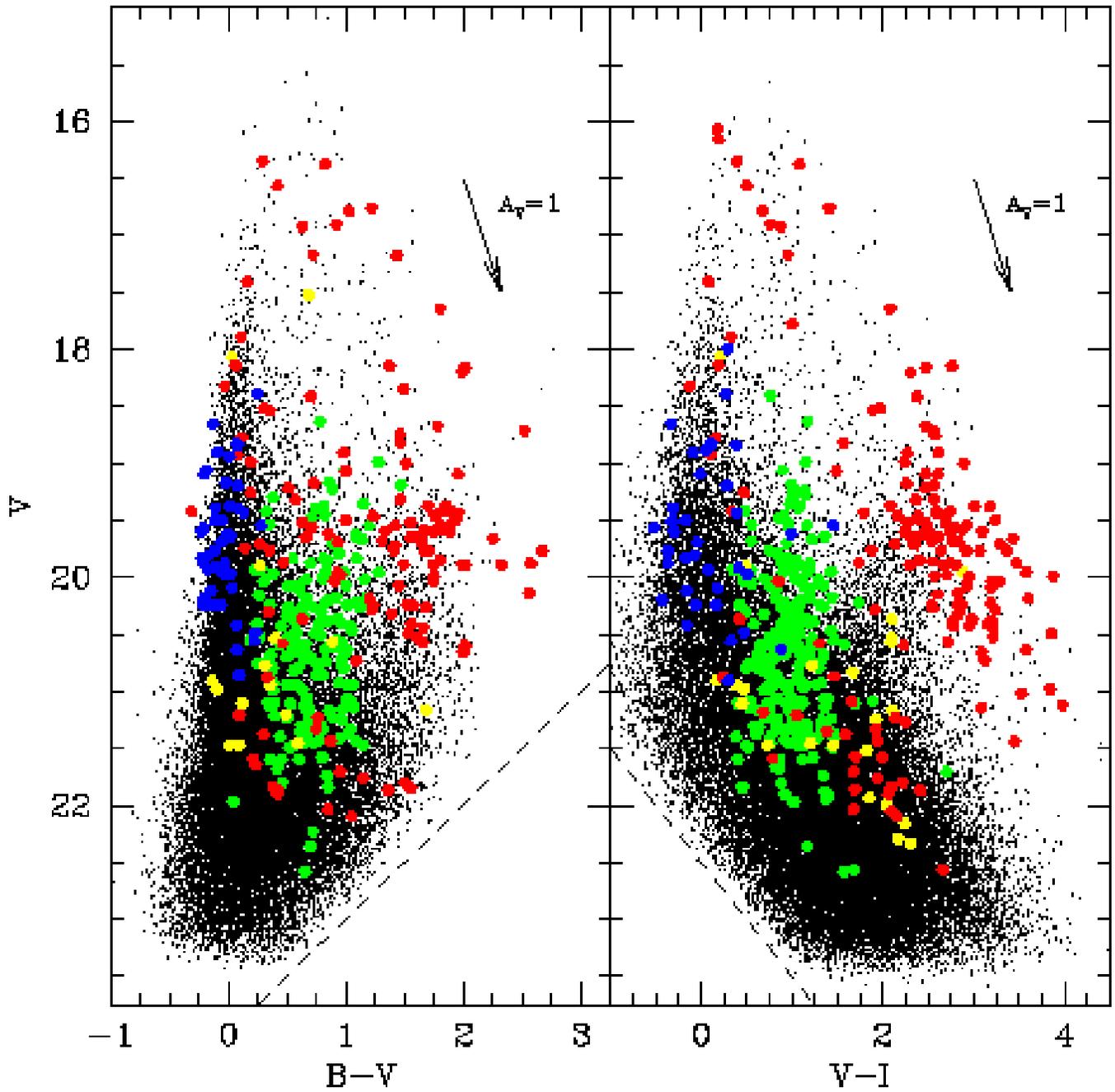}
\caption{Color-magnitude diagrams indicating the location of the
variables. Stars from the catalog of \citet{ma00a} are plotted using small
black dots, while variables are indicated using solid circles of different
colors: blue for EBs, green for Cepheids, yellow for periodic variables, and
red for miscellaneous variables.}
\end{figure}

\clearpage

\begin{figure}
\plotone{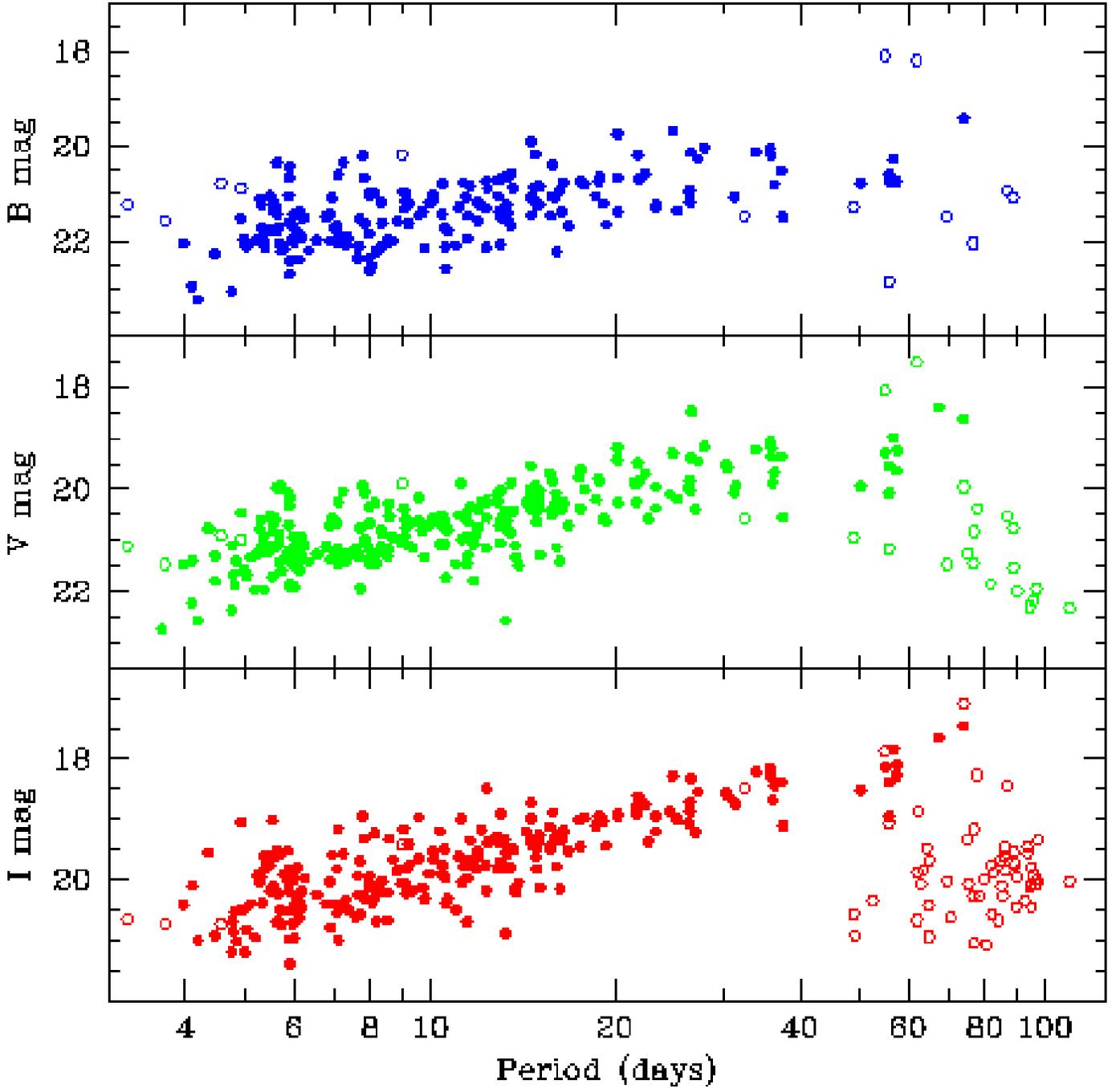}
\caption{$B$, $V$ and $I$-band Period-Luminosity relations for Cepheids (solid
circles) and ``periodic'' variables (open circles), constructed from the data
presented in Tables 1 and 3.}
\end{figure}

\clearpage

\begin{deluxetable}{lrrrccccccc}
\tablenum{1}
\tablecaption{DIRECT Cepheids}
\tablewidth{0pt}
\tabletypesize{\scriptsize}
\tablecolumns{11}
\tablehead{\colhead{Designation}&\multicolumn{2}{c}{R.A.\ \ (J2000.0)\ \ Dec.}&\colhead{P (d)}& \colhead{V}&\colhead{I}
&\colhead{B}& \colhead{$\sigma_V$}&\colhead{$\sigma_I$}&\colhead{$\sigma_B$}&\colhead{Amp.}}
\startdata
D33J013417.3+304505.3 & 01:34:17.35 & 30:45:05.25 &  3.66 & 22.74 &  \nd  &  \nd  & 0.14 &  \nd &  \nd & 0.79 \\ %A06715
D33J013341.2+302856.4 & 01:33:41.24 & 30:28:56.39 &  3.97 & 21.47 & 20.42 & 22.04 & 0.09 & 0.13 & 0.33 & 0.27 \\ %B24534
D33J013353.2+304420.4 & 01:33:53.18 & 30:44:20.37 &  4.10 & 22.23 &  \nd  & 22.95 & 0.19 &  \nd & 0.22 & 0.49 \\ %A20360
D33J013413.1+304210.5 & 01:34:13.12 & 30:42:10.55 &  4.11 & 21.39 & 20.10 &  \nd  & 0.11 & 0.10 &  \nd & 0.29 \\ %A09192
D33J013349.8+304737.0 & 01:33:49.80 & 30:47:37.03 &  4.19 & 22.58 & 21.01 & 23.23 & 0.21 & 0.23 & 0.20 & 0.56 \\ %A22123
D33J013336.9+303709.3 & 01:33:36.86 & 30:37:09.34 &  4.36 & 20.77 & 19.55 &  \nd  & 0.07 & 0.05 &  \nd & 0.21 \\ %B27000
D33J013351.9+304658.7 & 01:33:51.94 & 30:46:58.69 &  4.39 & 20.79 &  \nd  &  \nd  & 0.08 &  \nd &  \nd & 0.20 \\ %A20877
D33J013402.6+303944.5 & 01:34:02.60 & 30:39:44.53 &  4.47 & 21.30 &  \nd  &  \nd  & 0.09 &  \nd &  \nd & 0.33 \\
\enddata
\end{deluxetable}

\clearpage

\begin{deluxetable}{lrrrcccccccccc}
\tablenum{2}
\tablecaption{DIRECT Eclipsing Binaries}
\tablewidth{0pt}
\tabletypesize{\scriptsize}
\tablecolumns{14}
\tablehead{\colhead{Designation}&\colhead{R.A.}&\colhead{Dec.}&\colhead{Per.}& \colhead{V}&\colhead{I} &
\colhead{B}& \colhead{$\sigma_V$}&\colhead{$\sigma_I$}&\colhead{$\sigma_B$} & \colhead{$R_1$} & \colhead{$R_2$} &
\colhead{$i$} & \colhead{$e$} \\
\colhead{} & \multicolumn{2}{c}{(J2000.0)} & \colhead{(d)} & \colhead{} & \colhead{} & \colhead{} & 
\colhead{} & \colhead{} & \colhead{} & \colhead{} & \colhead{} & \colhead{($^\circ$)} & \colhead{}}
\startdata
\hspace{-0.5in}D33J013422.1+304710.4 & 01:34:22.08 & 30:47:10.39 &  1.0530 & 19.93 & 19.85 & 19.76 & 0.04 & 0.07 & 0.02 & 0.63 & 0.36 & 63.31 & 0.03 \\ %A04148
\hspace{-0.5in}D33J013332.0+303839.2 & 01:33:31.99 & 30:38:39.23 &  1.6691 & 19.64 &  \nd  & 19.66 & 0.06 &  \nd & 0.13 & 0.63 & 0.36 & 73.09 & 0.03 \\ %B30054
\hspace{-0.5in}D33J013418.0+303806.6 & 01:34:18.03 & 30:38:06.61 &  1.8717 & 19.88 &  \nd  & 19.85 & 0.06 &  \nd & 0.12 & 0.48 & 0.46 & 66.10 & 0.01 \\ %B01252
\hspace{-0.5in}D33J013415.0+303431.5 & 01:34:15.00 & 30:34:31.48 &  1.8773 & 19.92 & 19.51 & 19.81 & 0.04 & 0.07 & 0.04 & 0.65 & 0.35 & 53.97 & 0.03 \\ %B03186
\hspace{-0.5in}D33J013347.4+303249.0 & 01:33:47.44 & 30:32:49.02 &  1.8970 & 20.56 & 20.23 & 20.78 & 0.07 & 0.15 & 0.25 & 0.41 & 0.32 & 79.63 & 0.00 \\ %B20379
\hspace{-0.5in}D33J013354.4+304401.7 & 01:33:54.45 & 30:44:01.68 &  1.8989 & 20.64 & 19.76 & 20.71 & 0.08 & 0.12 & 0.03 & 0.63 & 0.37 & 65.28 & 0.00 \\ %A19727
\hspace{-0.5in}D33J013337.6+303803.1 & 01:33:37.59 & 30:38:03.08 &  1.9020 & 20.86 &  \nd  & 20.95 & 0.10 &  \nd & 0.25 & 0.44 & 0.35 & 85.99 & 0.02 \\ %B26488
\hspace{-0.5in}D33J013350.2+303317.8 & 01:33:50.24 & 30:33:17.82 &  2.0971 & 20.24 & 20.08 & 20.11 & 0.06 & 0.10 & 0.20 & 0.38 & 0.35 & 80.53 & 0.01 \\
\enddata
\end{deluxetable}

\clearpage

\begin{deluxetable}{lrrrcccccc}
\tablenum{3}
\tablewidth{0pt}
\tablecolumns{10}
\tablecaption{DIRECT periodic variables}
\tabletypesize{\footnotesize}
\tablehead{\colhead{Designation}&\multicolumn{2}{c}{R.A.\ \ (J2000.0)\ \ Dec.}&\colhead{P (d)}&\colhead{V}&\colhead{I}&\colhead{B}& \colhead{$\sigma_V$}&\colhead{$\sigma_I$}&\colhead{$\sigma_B$}}
\startdata
D33J013420.2+303400.7 & 01:34:20.18 & 30:34:00.73 &   3.2 & 21.11 & 20.66 & 21.23 & 0.17 & 0.14 & 0.13 \\ %B00162
D33J013334.2+303058.0 & 01:33:34.24 & 30:30:58.04 &   3.7 & 21.47 & 20.74 & 21.56 & 0.23 & 0.25 & 0.24 \\ %B28936
D33J013416.9+303454.6 & 01:34:16.87 & 30:34:54.58 &   4.6 & 20.91 & 20.74 & 20.77 & 0.13 & 0.19 & 0.08 \\ %B02052
D33J013339.7+302942.9 & 01:33:39.68 & 30:29:42.94 &   4.9 & 20.98 & 20.51 & 20.88 & 0.13 & 0.14 & 0.08 \\ %B25513
D33J013341.8+303452.1 & 01:33:41.83 & 30:34:52.06 &   6.0 & 21.20 &  \nd  & 21.69 & 0.23 &  \nd & 0.13 \\ %B23868
D33J013333.5+303320.5 & 01:33:33.50 & 30:33:20.49 &   9.0 & 19.90 & 19.42 & 20.17 & 0.13 & 0.12 & 0.13 \\ %B29316
D33J013409.8+303814.5 & 01:34:09.84 & 30:38:14.46 &  32.5 & 20.57 & 18.48 & 21.46 & 0.22 & 0.25 & 0.08 \\ %B06067
D33J013331.5+303255.0 & 01:33:31.50 & 30:32:55.03 &  49.0 & 20.94 & 20.56 & 21.29 & 0.27 & 0.49 & 0.10 \\
\enddata
\end{deluxetable}

\clearpage

\begin{deluxetable}{lrrrcccccc}
\tablecaption{DIRECT miscellaneous variables}
\tablenum{4}
\tablewidth{0pt}
\tablecolumns{10}
\tabletypesize{\footnotesize}
\tablehead{\colhead{Designation}&\multicolumn{2}{c}{R.A.\ \ (J2000.0)\ \ Dec.}&\colhead{$J_S$}&\colhead{V}&\colhead{I}& \colhead{B}& \colhead{$\sigma_V$}&\colhead{$\sigma_I$}&\colhead{$\sigma_B$}}
\startdata
D33J013331.1+303502.4 & 01:33:31.10 & 30:35:02.43 &  1.90 & 20.03 & 16.83 & 21.77 & 0.13 & 0.16 & 0.11 \\ %B30755
D33J013331.5+303410.2 & 01:33:31.51 & 30:34:10.17 &  1.40 & 19.43 & 17.64 & 21.39 & 0.08 & 0.06 & 0.07 \\ %B30554
D33J013333.2+303505.7 & 01:33:33.16 & 30:35:05.75 &  1.90 & 17.41 & 17.33 & 17.57 & 0.04 & 0.03 & 0.04 \\ %B29463
D33J013333.4+303146.7 & 01:33:33.35 & 30:31:46.70 &  2.26 & 20.64 & 17.06 &  \nd  & 0.23 & 0.06 &  \nd \\ %B29500
D33J013333.4+303350.7 & 01:33:33.41 & 30:33:50.73 &  7.82 & 18.91 & 16.30 & 22.40 & 0.46 & 0.24 & 0.35 \\ %B29352
D33J013333.5+303149.2 & 01:33:33.52 & 30:31:49.19 &  1.98 & 19.69 & 16.92 & 20.62 & 0.12 & 0.07 & 0.04 \\ %B29383
D33J013333.9+303402.6 & 01:33:33.85 & 30:34:02.57 &  2.99 & 19.42 & 16.83 & 21.32 & 0.21 & 0.09 & 0.06 \\ %B29056
D33J013334.8+303211.4 & 01:33:34.84 & 30:32:11.36 &  1.31 & 18.77 & 18.61 & 18.89 & 0.05 & 0.05 & 0.04 \\
\enddata
\end{deluxetable}

\clearpage

\begin{deluxetable}{ccc}
\tablenum{5}
\tablecaption{Photometric data - Cepheid variables}
\tablewidth{0pt}
\tabletypesize{\normalsize}
\tablecolumns{3}
\tablehead{\colhead{HJD}&\colhead{Magnitude}&\colhead{$\sigma$}\\
\colhead{-2450000} & \colhead{} & \colhead{}}
\startdata
\multicolumn{3}{l}{\bf D33J013417.3+304505.3 - V} \\
332.7556 & 23.042 & 0.599 \\
332.7855 & 22.943 & 0.242 \\
332.9925 & 22.285 & 0.148 \\
333.8915 & 22.514 & 0.164 \\
333.8982 & 22.256 & 0.153 \\
334.8091 & 22.925 & 0.201 \\
334.8157 & 23.057 & 0.240 \\
334.8274 & 22.847 & 0.166 \\
348.7933 & 22.682 & 0.252 \\
348.8074 & 22.675 & 0.246 \\
348.8125 & 22.357 & 0.178 \\
348.8243 & 22.610 & 0.165 \\
348.9463 & 22.587 & 0.141 \\
348.9580 & 22.913 & 0.174 \\
348.9697 & 22.911 & 0.206 \\
348.9815 & 22.769 & 0.157 \\
\multicolumn{3}{l}{\it Continues in electronic form} \\
\enddata
\end{deluxetable}

%\clearpage

\begin{deluxetable}{ccc}
\tablenum{6}
\tablecaption{Photometric data - Eclipsing binaries}
\tablewidth{0pt}
\tabletypesize{\normalsize}
\tablecolumns{3}
\tablehead{\colhead{HJD}&\colhead{Magnitude}&\colhead{$\sigma$}\\
\colhead{-2450000} & \colhead{} & \colhead{}}
\startdata
\multicolumn{3}{l}{\bf D33J013422.1+304710.4 - V} \\
332.7272 & 19.881 & 0.038 \\
332.7556 & 20.013 & 0.040 \\
332.7787 & 19.942 & 0.032 \\
332.7855 & 19.984 & 0.030 \\
332.9925 & 20.152 & 0.038 \\
333.8915 & 20.036 & 0.029 \\
333.8982 & 20.081 & 0.030 \\
334.8091 & 19.937 & 0.029 \\
334.8157 & 19.916 & 0.032 \\
334.8274 & 19.952 & 0.026 \\
348.7933 & 20.096 & 0.037 \\
348.8074 & 20.129 & 0.050 \\
348.8125 & 20.084 & 0.038 \\
348.8243 & 20.089 & 0.040 \\
348.9463 & 19.918 & 0.040 \\
348.9580 & 19.928 & 0.028 \\
\multicolumn{3}{l}{\it Continues in electronic form} \\
\enddata
\end{deluxetable}

%\clearpage

\begin{deluxetable}{ccc}
\tablenum{7}
\tablecaption{Photometric data - Periodic variables}
\tablewidth{0pt}
\tabletypesize{\normalsize}
\tablecolumns{3}
\tablehead{\colhead{HJD}&\colhead{Magnitude}&\colhead{$\sigma$}\\
\colhead{-2450000} & \colhead{} & \colhead{}}
\startdata
\multicolumn{3}{l}{\bf D33J013420.2+303400.7 - V}\\
332.8749 & 20.863 & 0.054 \\
332.8866 & 20.836 & 0.053 \\
332.8984 & 20.912 & 0.060 \\
333.9348 & 21.164 & 0.072 \\
333.9466 & 21.176 & 0.060 \\
334.7195 & 21.314 & 0.105 \\
334.7367 & 21.354 & 0.116 \\
334.9713 & 21.233 & 0.070 \\
334.9830 & 21.252 & 0.064 \\
334.9948 & 21.381 & 0.069 \\
348.8472 & 20.865 & 0.070 \\
348.8589 & 20.874 & 0.067 \\
348.8707 & 20.787 & 0.063 \\
348.8972 & 20.874 & 0.063 \\
349.0038 & 20.919 & 0.063 \\
\multicolumn{3}{l}{\it Continues in electronic form} \\
\enddata
\end{deluxetable}

%\clearpage

\begin{deluxetable}{ccc}
\tablenum{8}
\tablecaption{Photometric data - Miscellaneous variables}
\tablewidth{0pt}
\tabletypesize{\normalsize}
\tablecolumns{3}
\tablehead{\colhead{HJD}&\colhead{Magnitude}&\colhead{$\sigma$}\\
\colhead{-2450000} & \colhead{} & \colhead{}}
\startdata
\multicolumn{3}{l}{\bf D33J013331.1+303502.4 - V}\\
332.8749 & 19.741 & 0.033 \\
332.8866 & 19.726 & 0.033 \\
332.8984 & 19.743 & 0.029 \\
333.9348 & 19.735 & 0.032 \\
333.9466 & 19.742 & 0.033 \\
334.7195 & 19.825 & 0.039 \\
334.7367 & 19.767 & 0.037 \\
334.9713 & 19.748 & 0.035 \\
334.9830 & 19.707 & 0.029 \\
334.9948 & 19.689 & 0.029 \\
348.8472 & 19.892 & 0.037 \\
348.8589 & 19.916 & 0.043 \\
348.8707 & 19.932 & 0.046 \\
348.8972 & 19.853 & 0.037 \\
349.0038 & 19.838 & 0.033 \\
349.9414 & 19.769 & 0.067 \\
\multicolumn{3}{l}{\it Continues in electronic form} \\
\enddata
\end{deluxetable}

\clearpage

\begin{deluxetable}{lclllccccc}
\tablenum{9}
\tablecaption{M33 Classical variables}
\tabletypesize{\footnotesize}
\tablewidth{0pt}
\tablecolumns{9}
\tablehead{
\multicolumn{2}{c}{Designation} & \colhead{R.A.} & \colhead{Dec.} & \multicolumn{2}{c}{Type} & \multicolumn{2}{c}
{Per (d)}& \colhead{Notes} \\
\colhead{Classical} & \colhead{DIRECT} & \multicolumn{2}{c}{(J2000.0)} & \colhead{Cl.} & \colhead{DI.} & \colhead{Cl.}&
\colhead{DI.} & \colhead{}}
\startdata
Du22-V01 & D33J013409.8+303814.5 & 01:34:09.8 & 30:38:15 & V & C &  \nd  &  32.55 &     \\
Du22-V02 & D33J013418.4+303837.1 & 01:34:18.4 & 30:38:37 & V & N &  \nd  &   \nd  &     \\
Hu26-V03 &          \nd          & 01:33:25.9 & 30:48:34 & C &\nd& 41.68 &   \nd  &     \\
Hu26-V04 & D33J013423.6+304204.6 & 01:34:23.6 & 30:42:05 & C & C & 27.37 &  27.28 &  1  \\
Hu26-V05 & D33J013404.9+304022.6 & 01:34:04.9 & 30:40:23 & C & C & 14.60 &  14.63 &     \\
Hu26-V06 &          \nd          & 01:34:45.1 & 30:49:25 & V &\nd&  \nd  &   \nd        \\
Hu26-V07 & D33J013411.3+304156.3 & 01:34:11.3 & 30:41:56 & C & C & 26.56 &  26.56 &     \\
Hu26-V08 & D33J013355.1+302957.4 & 01:33:55.1 & 30:29:57 & E & E &  \nd  &  5.095 &     \\
\enddata
\tablerefs{Du22: \citet{du22}; Hu26: \citet{hu26}; HS53: \citet{hs53}; VHK75: 
\citet{vhk75}; SC83: \citet{sc83}}

\tablecomments{Key to types: C: Cepheid; E: Eclipsing binary; P: periodic
variable; V: miscellaneous variable; N: does not display variability in the
DIRECT database; ($\dots$): object outside the boundaries of the catalog of
\citet{ma00a}, coordinates obtained from the literature or the digitized POSS
plate of M33.}

\tablecomments{[1]: possibly mis-identified with nearby star in original
finding chart; [2]: variable located in DIRECT field C \citep{st01}; [3]:
saturated in DIRECT images; [4]: same object as Hu26-V04; [5]: very faint,
below detection threshold in DIRECT images; [6]: very faint, below detection
threshold in DIRECT images; [7] possibly mis-identified with nearby star in
original finding chart; [8]: same object as Hu26-V30}
\end{deluxetable}

\clearpage

\begin{deluxetable}{lcllcccc}
\tablenum{10}
\tablecaption{Variables in common with \citet{kmw87}}
\tabletypesize{\small}
\tablewidth{0pt}
\tablecolumns{8}
\tablehead{
\multicolumn{2}{c}{Designation} & \colhead{R.A.} & \colhead{Dec.} & \colhead{DIR.} & \multicolumn{2}{c}
{Per. (d)} & \colhead{Notes} \\
\colhead{KMW} & \colhead{DIRECT} & \multicolumn{2}{c}{(J2000.0)} & \colhead{type} & \colhead{K}& \colhead{D} & \colhead{}}
\startdata
\multicolumn{8}{l}{\it Objects classified as Cepheids or periodic (Tables III \& IV)} \\
P11721 & D33J013412.2+304641.4 & 01:34:12.4 & 30:46:41.4 & C & 35.90 & 35.86 &     \\
P13013 & D33J013409.6+304536.0 & 01:34:09.6 & 30:45:36.0 & C & 25.26 & 25.29 &     \\
P16673 & D33J013402.8+304145.7 & 01:34:02.8 & 30:41:45.7 & C & 25.20 & 13.04 &     \\
P22263 & D33J013347.2+303536.2 & 01:33:47.2 & 30:35:36.2 & C & 26.50 & 26.50 &     \\
P23601 & D33J013332.9+303548.4 & 01:33:32.9 & 30:35:48.5 & C & 30.42 & 30.29 &     \\ 
P28325 & D33J013331.4+303016.3 & 01:33:31.4 & 30:30:16.3 & C & 26.91 & 26.98 &     \\ 
P28834 & D33J013331.3+302946.9 & 01:33:31.3 & 30:29:46.9 & C & 17.58 & 17.54 &     \\
\enddata
\tablecomments{Key to DIRECT types: C: Cepheid; P: periodic variable; V: miscellaneous variable; N:
non-variable star present in the catalog of \citet{ma00a}; ; $(\dots)$: no star present in the catalog of
\citet{ma00a} at this position.}
\tablecomments{[1]: very faint, below detection threshold in DIRECT images. [2]: classified by KMW87 as a W Vir
candidate.}
\end{deluxetable}
\end{document}